\documentclass[%
reprint, 
superscriptaddress, 
showpacs,preprintnumbers,
amsmath,amssymb,
aps,
prl,
]{revtex4-2}

\usepackage{graphicx}
\usepackage{dcolumn}
\usepackage{bm}
\usepackage{siunitx}
\usepackage{xcolor}
\usepackage{soul}
\usepackage[T1]{fontenc}
\usepackage{lmodern}

\begin{document}

\preprint{APS/123-QED}

\title{What determines the breakup length of a jet?}

\author{Stefan Kooij}
\affiliation{Van der Waals-Zeeman Institute, University of Amsterdam, Science Park 904, Amsterdam, Netherlands
}%
\author{Daniel T. A. Jordan}
  \affiliation{ 
Institute of Physics, University of Amsterdam, Science Park 904, 1098 XH Amsterdam, The Netherlands
}%
  \author{Cees J.M. van Rijn}
\affiliation{Van der Waals-Zeeman Institute, University of Amsterdam, Science Park 904, Amsterdam, Netherlands
}%
  \author{Neil Ribe}
\affiliation{Laboratoire FAST, Universit\'{e} Paris-Saclay, CNRS, 91405 Orsay, France
}%
\author{Daniel Bonn}
\affiliation{Van der Waals-Zeeman Institute, University of Amsterdam, Science Park 904, Amsterdam, Netherlands
}%

\date{\today}

\begin{abstract}
The breakup of a capillary jet into drops is believed to be governed by initial disturbances on the surface of the jet that grow exponentially. The disturbances are often assumed to be due to external sources of noise, to turbulence, or to imperfections of the nozzle. However, even in conditions where external perturbations are minimal, the jet’s length cannot grow indefinitely, suggesting that its fragmentation cannot be entirely attributed to such factors. Here we show that the initial disturbances are thermal capillary waves. By extrapolating the observed growth of the instability back in time, we demonstrate that the initiating disturbances must be of the order of an Ångström, consistent with fluctuations induced by thermal noise. Further, by performing many experiments with different nozzles, we find no significant variation in breakup length linked to nozzle type, shape or inner roughness. By systematically varying the jet diameter and velocity, and the surface tension, viscosity, and density of the fluid, we validate our thermal disturbance model over four orders of magnitude in jet length; seven orders of magnitude if simulations of nanojets are included.

\end{abstract}

\keywords{Suggested keywords}

\maketitle


The breakup of laminar liquid jets is a phenomenon of fundamental importance in both natural and industrial processes, with applications ranging from inkjet printing and DNA sampling to food processing and drug delivery \citep{lohse2022fundamental, zhang2023improved, d2023fluorescence, Pharm_chemical_processes, food-processing, eggers2008physics}. The seminal work of Lord Rayleigh showed that breakup of an axisymmetric jet is due to small initial disturbances that grow exponentially \citep{rayleigh1879capillary,weber1931zerfall,phinney1972stability,donnelly1966experiments,rutland1971non,eggers2008physics}. The initial perturbations are often assumed to be due to environmental sources of noise such as vibrations, or to imperfections of the liquid flow through the orifice or capillary. Savart noticed in 1833 that the intact length of a jet may be increased\textit{``by suitable insulation of the reservoir from tremors"}, but added that \textit{``Nevertheless it does not appear to be possible to carry the prolongation very far."}\citep{rayleigh1879capillary}. Savart's observation has been supported by later studies, which suggest that there is always some level of residual disturbances that can lead to breakup. However, the origin of these disturbances remains unclear \citep{phinney1972stability}. A recent suggestion is that the disturbances may be due to previous breakup events themselves, leading to a self-sustaining `natural' breakup length for the jet \citep{ganan2021natural}. Evidently, the breakup of liquid jets remains an outstanding scientific problem nearly 200 years after Savart's pioneering work.

\begin{figure}[!b]
\centering
\includegraphics[width = 0.45\textwidth]{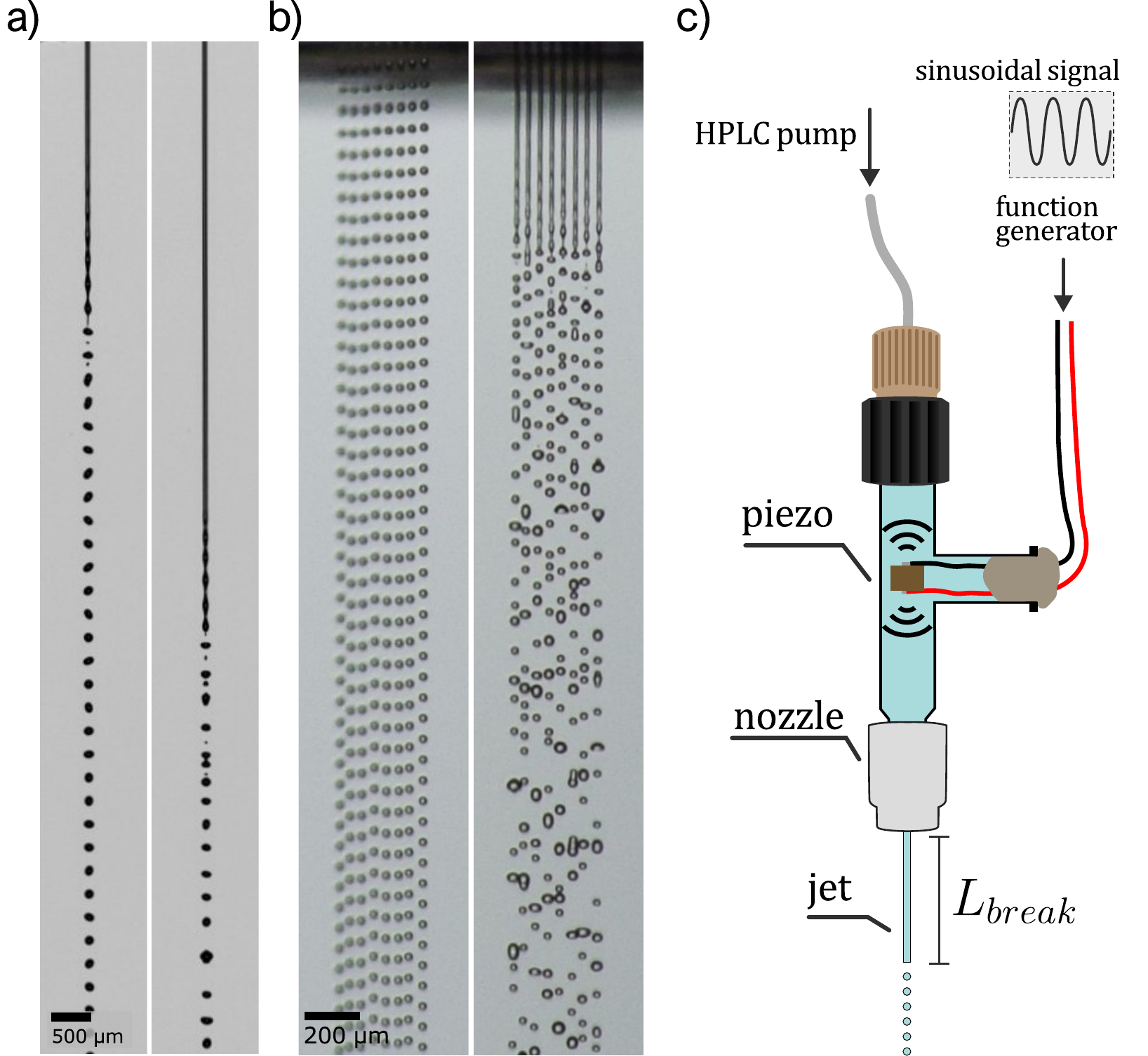}
\caption{Perturbed versus unperturbed water jets. a) A single \SI{64}{\micro \meter} diameter jet at \SI{2}{\bar} operating pressure, without perturbations (right) and perturbed at a frequency \SI{28.2}{\kilo \hertz} (left). b) Eight \SI{16}{\micro \meter} diameter jets, unperturbed (right) and perturbed (left). A significant variability in droplet sizes occurs for the unperturbed jets, and increases with downstream distance due to coalescence. c) Jets are perturbed using a submerged piezoelectric transducer.   \footnotesize }
\label{fig:pictures_jets_perturbed_unperturbed}
\end{figure}

Since the natural breakup of a jet is not perfectly regular, significant variability in breakup lengths and droplet sizes is typically observed (Fig.~\ref{fig:pictures_jets_perturbed_unperturbed}). Because this variability can negatively affect performance in applications such as ink-jet printing, it is crucial to understand the mechanisms that govern jet breakup. Here we investigate the origin of the initial disturbances and find, contrary to prevailing assumptions, that they are not due to nozzle imperfections, environmental noise, or noise from the breakup itself. Instead, we propose that the disturbances are due to thermal capillary waves, i.e. thermal fluctuations, recently shown in simulations to influence breakup patterns \citep{zhao2020dynamics, barker2023fluctuating}. Using high-speed imaging, we track the growth of visible disturbances on the jet surface and extrapolate their exponential growth back to $t=0$, thereby obtaining an estimate of the magnitude of the initial perturbations. The result is on the order of 1 \r{A}, which is typical of the amplitude of thermal capillary waves or thermal surface roughness. Combining our experimental results with MD simulations of nanojets for which thermal noise is important \citep{moseler2000formation}, we find that a simple model of breakup triggered by thermally-driven disturbances accurately predicts jet lengths over more than seven orders of magnitude, from nanoscale- to macroscopic jets.

\begin{figure}[!b]
\centering
\includegraphics[width = 0.45\textwidth]{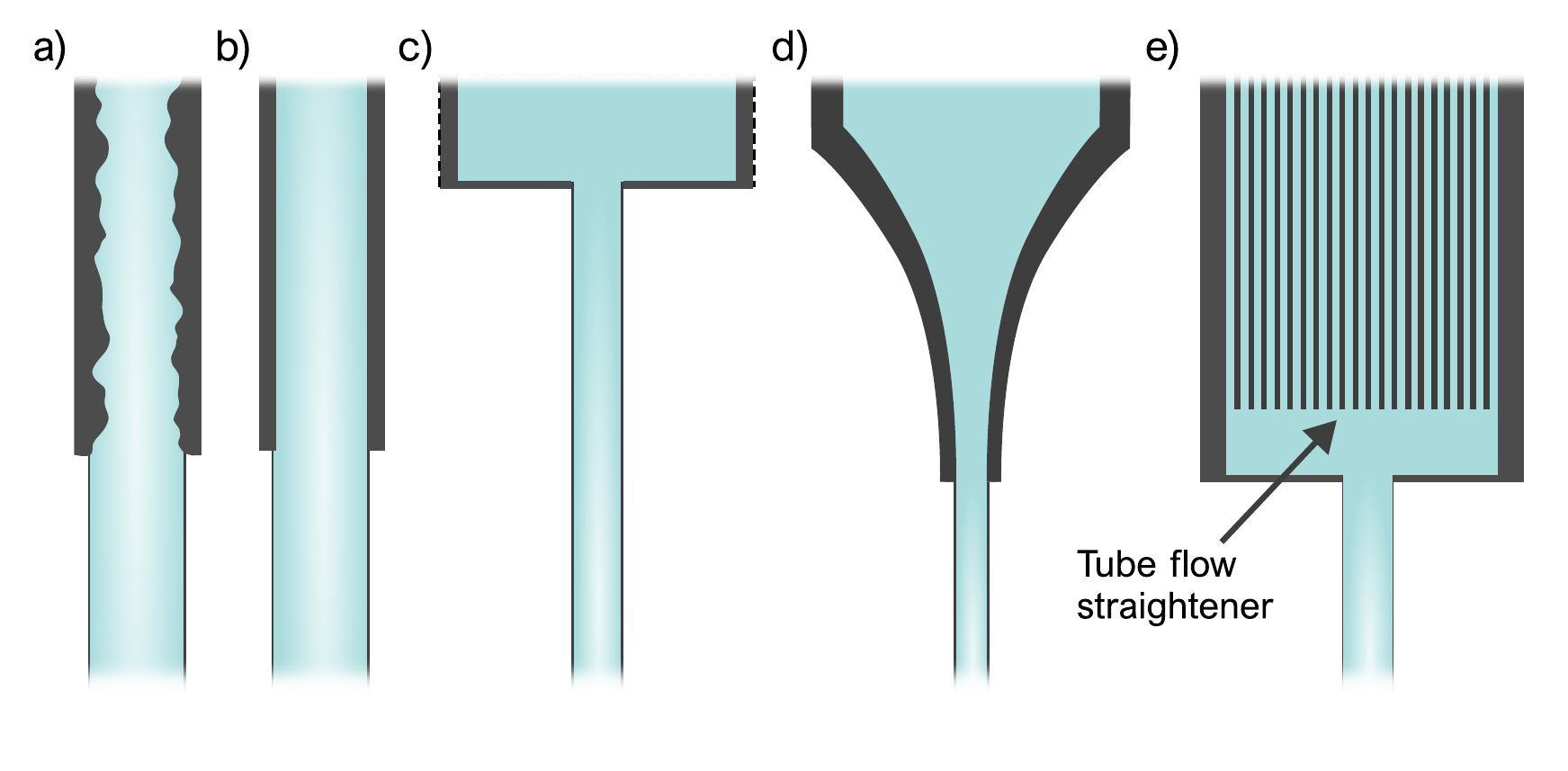}
\caption{ Jet nozzle geometries. a) Stainless steel needles with a flat end (Techcon) and relatively rough inner surfaces. b) Glass capillaries, with very smooth inner surfaces, c) Flat-plate nozzles of variable quality, ranging from thin metallic plates with holes of uneven perimeter to micro-fabricated silicon spray nozzles (Medspray) with smooth holes. d) Smooth stainless steel jet nozzles with tapered profiles (Schlick). e) Our in-house design: jet orifices cut by EDM in thin metallic plates attached to a flow straightener made using plastic straws, sponges and meshes to suppress turbulence.  \footnotesize }
\label{fig:nozzle_geometries}
\end{figure}

\begin{figure}[!t]
\centering
\includegraphics[width = 0.45\textwidth]{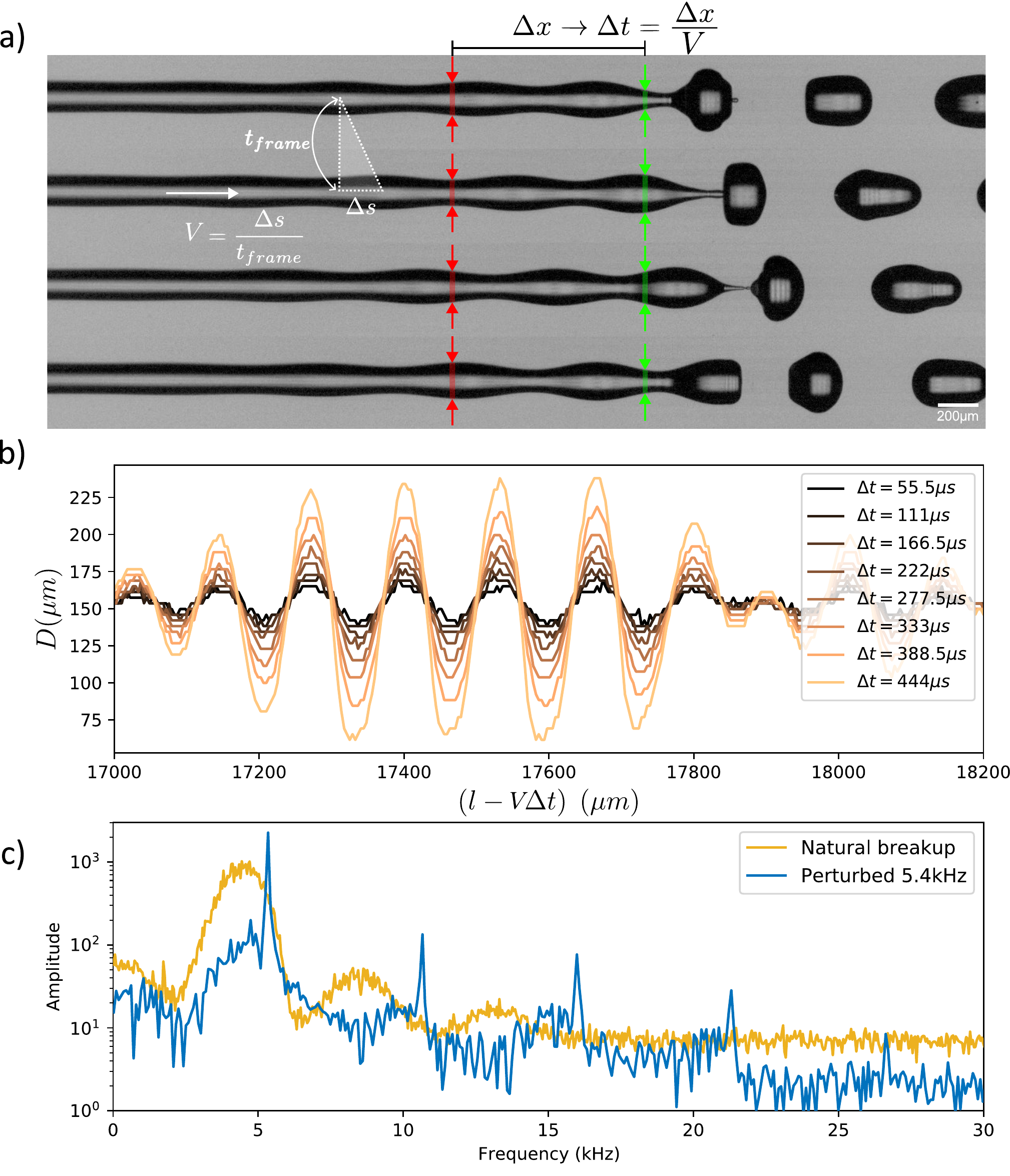}
\caption{Analysis of jet undulations $D=$ \SI{150}{\micro \meter}, $\Delta t =$ \SI{55.5}{\micro \second} a) Four consecutive images of an unperturbed water jet destabilizing. Using image analysis, the width of the jet is determined at different positions for a sequence of high-speed images. For illustrative purposes two positions are indicated with red and green arrows. Each point yields a sinusoidal pattern when the time shift $\Delta t = \Delta x/V$ between measurements is taken into account. b) Evolution of the spatially varying jet width $D$, corrected for the distance traveled. c) Fourier analysis of the jet's surface for natural breakup (orange) and for a jet perturbed at a frequency \SI{5.4}{\kilo \hertz} (blue).    \footnotesize }
\label{fig:wave_pattern_analysis}
\end{figure}

In our experiments we use a variety of nozzle geometries (Fig.~\ref{fig:nozzle_geometries}). The hole diameter was varied by a factor of $10^3$, from \SI{4}{\milli \meter} down to \SI{4}{\micro \meter}. We varied the viscosity $\mu$ from 1-15\;\SI{}{\milli \pascal \second} using water-glycerol solutions. The surface tension $\gamma$ was varied by using both water-ethanol solutions (22-72\;\SI{}{\milli \newton \per \meter}) and the eutectic liquid Galinstan (\SI{718}{\milli \newton \per \meter}). The density $\rho$ was also varied using the aforementioned liquid mixtures as well as Galinstan (density \SI{6.44}{\gram \per \cubic \centi \meter}). The Galinstan jet experiment was performed under vacuum  to prevent the formation of an oxide layer. To pressurize the liquid we employed several techniques depending on the nozzle and the experiment. These included an HPLC pump (Waters 515), a pressure tank, a hand-held syringe, or simply the tap water installation. Flow rates were determined using image analysis or pressure sensor data, or by collecting and weighing the ejected liquid over a set time interval. We measured the breakup length in 158 separate experiments using either a high speed camera (Phantom TMX 7510) or a regular camera with a fast flashlight (Vela One). We used simple image analysis (Python) to determine the jet lengths and the modulations of the jet thickness  (Fig.~\ref{fig:wave_pattern_analysis} (a). We excluded turbulent jets because the onset of turbulence quickly reduces the jet length \citep{phinney1972stability}). Because the breakup length naturally fluctuates, we estimated it by averaging over a large number of measurements.  Most of the experiments were conducted on an optical table, and no particular effort was made to isolate the experiment from external sources of noise such as the pumps or the fan of the high-speed camera.

\begin{figure}[!t]
\centering
\includegraphics[width = 0.45\textwidth]{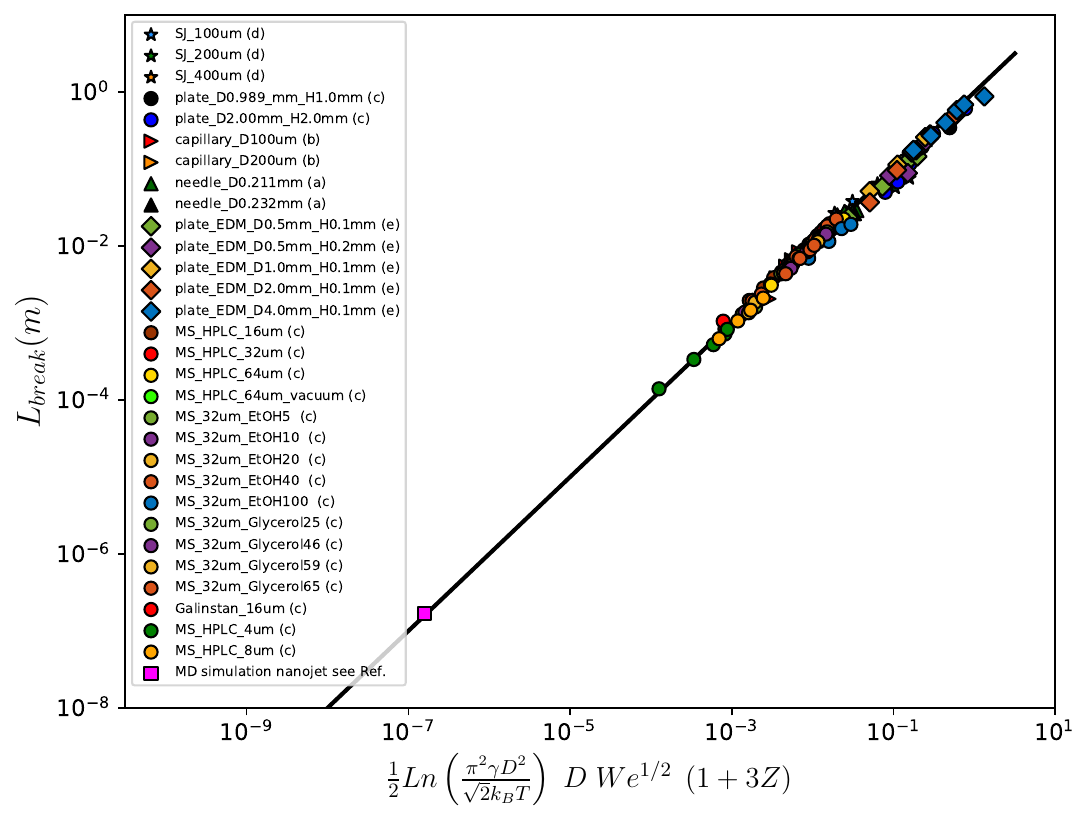}
\caption{Measured jet lengths $L_{break}$ for various nozzle geometries, liquid velocities, fluid properties and nozzle diameters versus the predicted value according to equation \eqref{eq:final_scaling}. The result from MD simulations of a nanojet is also included \citep{moseler2000formation}.   \footnotesize }
\label{fig:plot_jet_lengths}
\end{figure}

To examine the origin of the initial disturbances, we first focus on the prevailing idea that nozzle imperfections and environmental noise are responsible for the jet breakup. Experiments using a variety of nozzles with different geometries (Fig.~\ref{fig:nozzle_geometries}), wall roughnesses and liquid flow-profiles all collapse onto the theoretical prediction that we derive below (Fig.~\ref{fig:plot_jet_lengths}). Although the onset of turbulence is clearly influenced by nozzle quality, as long as the jet remains laminar it made no difference whether smooth, high-quality tapered nozzles, or capillaries with rough outlets were used. To study the effect of environmental sources of noise, we minimized acoustic coupling with the environment by placing the whole experiment on an optical table in an isolated plexiglass chamber,  using an external HPLC pump with a minimal noise level. The chamber was placed under vacuum to further minimize noise, and the pump was connected to the spray nozzle using a long and thin flexible plastic tube. Even at this level of isolation, we could not detect any increase in breakup length for several different liquid velocities. In short: although acoustic background noise can perturb a jet - noise was linked to variability in jet length long ago \citep{Savart_1833, Lafrance_noisy-setup} - we find that our experiments yield practically indistinguishable results with close to zero acoustic coupling. This suggests that background noise becomes unimportant below a certain threshold and reinforces the idea that the initial disturbances originate intrinsically.

How initial disturbances affect breakup becomes most apparent when the jet surface is actively perturbed. Figure \ref{fig:pictures_jets_perturbed_unperturbed}a shows perturbed (left) and unperturbed (right) jets issuing from a nozzle with \SI{64}{\micro \meter} diameter, and figure \ref{fig:pictures_jets_perturbed_unperturbed}b is similar but for eight adjacent jets issuing from holes with \SI{16}{\micro \meter} diameter.  In both cases, an immersed piezoelectric transducer was used to generate controlled disturbances on the jet surface. The impact of controlled perturbations is profound: not only do the excitations result in perfectly monodisperse droplets, but the jet length is also significantly reduced. Without active perturbations, the droplets resulting from breakup already vary in size, and this variability only increases downstream due to coalescence. 
The controlled application of initial perturbations shows how the breakup mechanism can be influenced when the origin of the disturbances is known.

The relationship between initial disturbances and jet breakup is described by Rayleigh's linearized perturbation theory \citep{rayleigh1879capillary}. Rayleigh showed that an axisymmetric jet is inherently unstable due to its out-of-equilibrium energy state, leading to the so-called Rayleigh-Plateau instability, in which small disturbances on the jet's surface grow exponentially. The radius $R(z,t)$ of the jet varies as 
\begin{equation}
    R(z,t) = R + \delta_k \cos{(kz)} \, e^{\Omega_k \, t} \; ,
\label{eq:Rayleigh}
\end{equation}
where $R$ is the initial radius of the jet and $k$ and $\delta_k$ are the wavenumber and amplitude of the perturbation, respectively. 
The growth rate
$\Omega_k$ is given by the well-known dispersion relation \citep{rayleigh1879capillary,weber1931zerfall,eggers2008physics} 
\begin{equation}
    \Omega_k = \sqrt{\frac{\gamma}{2\rho R^3}} \left( \sqrt{(x^2-x^4)+\frac{9}{2} \, Z^2 \, x^4} -3 \, Z \, x^2\right) \; ,
    \label{eq:dispersion} 
\end{equation}
where $x = k R$ and $Z = \mu/\sqrt{2\rho R\gamma}$ is the Ohnesorge number. The unstable modes have wavelengths larger than the jet's perimeter ($\, \lvert kR \rvert  < 1$) and the wavelength $\lambda_R$ and frequency of the fastest growing mode are called the Rayleigh wavelength and frequency. Even the smallest axisymmetric disturbance on the jet's surface is strongly amplified, ultimately leading to breakup after a time $T_k$ at which the perturbation's amplitude  $\delta_k \, e^{\Omega_k \, T_k} $ is of the order of the radius $R$. At constant jetting speed $V$, we can therefore estimate the breakup length $L_k$ of a perturbed jet to be
\begin{equation}
   L_k=V \cdot T_k = \frac{V}{\Omega_k} \, \ln{\left( \frac{R}{\delta_k} \right)} \; .
\label{eq:breaktime}
\end{equation}
The jet length is therefore a direct measure of the amplitude of the initial disturbance for a perturbation with a fixed wavelength. The logarithmic dependence in (\ref{eq:breaktime}) leads to a weak dependence of the breakup length on the perturbation level $\delta_k$, which may explain the difficulty in determining the origin of the perturbations. However, the change in breakup length becomes apparent if there are large differences in $\delta_k$, as there are between the unperturbed and perturbed jets in Fig.~\ref{fig:pictures_jets_perturbed_unperturbed}.

Figure \ref{fig:wave_pattern_analysis} (a) shows consecutive frames of a destabilizing jet. Using image analysis, the width of the jet can be determined at a given point in the lab frame  for the whole duration of the video, resulting in a sinusoidal wave pattern. By doing this for various distances from the nozzle we can resolve the evolution of the disturbances. An example is shown in Fig.~\ref{fig:wave_pattern_analysis}b,c. A Fourier transform of this pattern shows higher harmonics appearing close to the breakup point, i.e. smaller peaks at integer multiples of the main peak's frequency (Fig.~\ref{fig:wave_pattern_analysis}c).
These higher harmonics are probably due to the non-linearity of the final phase of the droplet pinch-off when the deformed jet shape is no longer sinusoidal \citep{Yuen_1968_non-linear_harmonics, Rutland_non-linear-harmonics}. Apart from the amplitude, the wave pattern itself does not change.  This rules out the idea that the pinch-off itself introduces perturbations that travel upstream \citep{gonzalez2021self}, since these would be noticeable in this analysis. Instead, our analysis confirms Rayleigh's original idea that initial disturbances determine the jet length through \eqref{eq:breaktime}. The Fourier spectrum of a perturbed jet is a series of sharp peaks, a direct consequence of the pure sinusoidal perturbation we applied. In contrast, the spectrum of an unperturbed jet is smooth, broad and peaked around the Rayleigh frequency. This confirms the assumption leading to \eqref{eq:scaling}, and supports the idea of a random (or noisy) origin of the initial disturbances.

We consider two possibilities for an intrinsic source of disturbances: they are caused either by the jet breakup itself, or by the intrinsic roughness of the liquid surface. Starting with the former, we note that the pinch-off of droplets - which ``kick" the jet at around the resonant frequency - might be the main source of irremovable disturbances, which could travel back up to the nozzle and induce another breakup event after a time $\Delta t \sim L_{break}/V$. Savart already demonstrated that when the impact of the stream of produced droplets is acoustically coupled with the nozzle, the breakup self-organizes to a steady mode \citep{donnelly1966experiments, gonzalez2021self}. As there is no such a steady breakup for the unperturbed jet, it seems unlikely that the breakup length is self-limited. Still, to further test this idea, we performed another experiment in which a jet fell onto a liquid bath just above the breakup point. The bath was then quickly lowered while the jet length was recorded with a high-speed camera. On the time scale of the breakup, the lowering of the liquid bath is slow, as the jet speed $\approx$\SI{10}{\meter \per \second}. If the breakup itself is a source of perturbations, one would expect to be able to temporarily extend the jet's length by increasing the distance between the liquid bath and the nozzle. However, we were unable to observe any increase or decrease of the jet length, demonstrating that the noise induced by the downstream breakup does not affect the jet length.

\begin{figure}[!b]
\centering
\includegraphics[width = 0.45\textwidth]{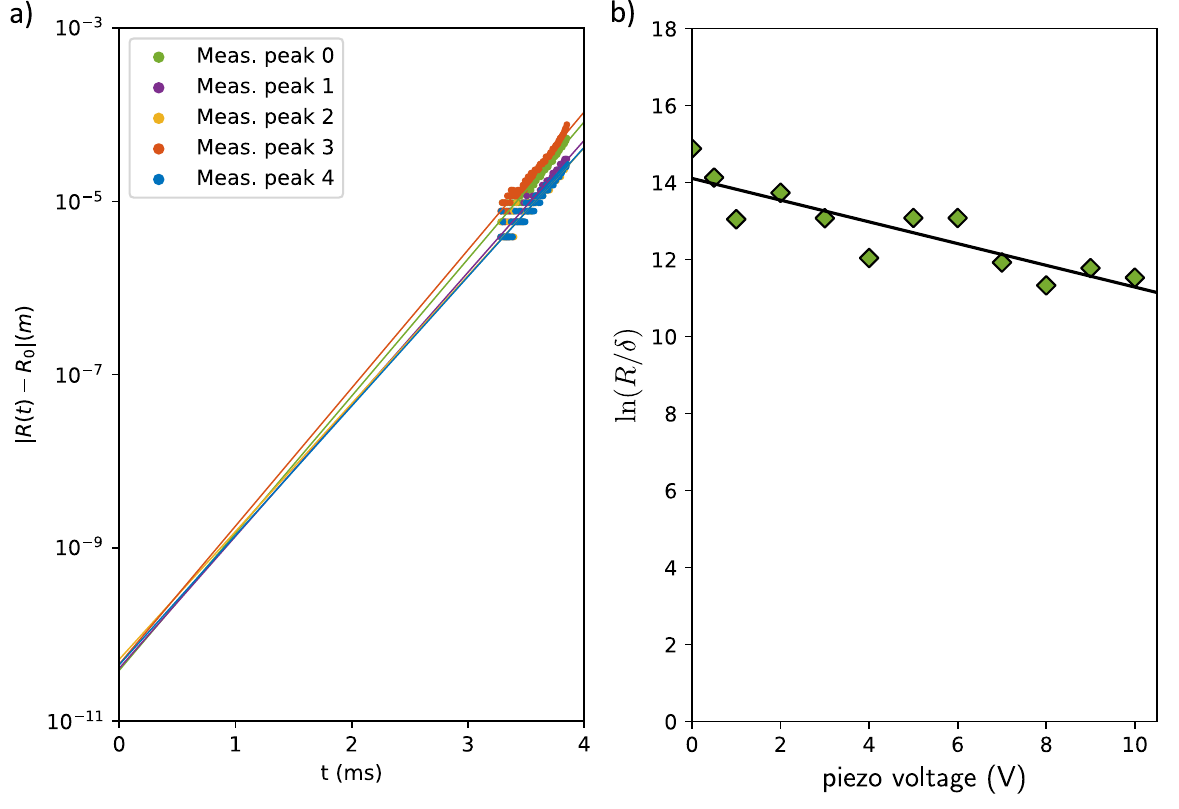}
\caption{Growing disturbances. a) Exponential fits of tracked thickness modulations on the surface of the jet, as in Fig.~\ref{fig:wave_pattern_analysis}. The intersection with the $y$-axis provides an estimate of the magnitude of the initial perturbation $\delta_0$ at the nozzle. b) Logarithmic factors derived from the exponential fits of the growth of the perturbations for different voltage levels of the piezoelectric transducer. \footnotesize }
\label{fig:exponental_fit}
\end{figure}

These experimental results do not support a role for external noise or a feedback mechanism close to breakup. Instead, they suggest that the breakup mechanism of an unperturbed jet is due to intrinsic perturbations close to the orifice. We therefore return to Rayleigh's theory to see what the origin of the initial perturbations might be. Using data like those shown in Fig. \ref{fig:wave_pattern_analysis}b, we can track the growth of the disturbances on the jet surface and  extrapolate the exponential growth back to $t=0$, i.e. to the origin of the jet at the nozzle. This allows one to estimate the magnitude of the initial perturbations through \eqref{eq:Rayleigh}. Figure \ref{fig:exponental_fit}(a) shows several examples of such exponential fits, all of which converge to an intersection with the $y$-axis $\delta_0 \simeq$ 1 \r{A}. Although this method is imprecise, as there is a large range of length scales not covered by the experiments, it still provides a reasonable estimation of $\delta_0$. Moreover, earlier work \citep{phinney1972stability} showed that the breakup length \eqref{eq:scaling} for silicone oils reached a plateau at $\ln{\left(R/\delta\right)} \sim 15$, which implied $\delta_0\approx$ 1 \r{A}. Though the authors did not address the origin of this plateau, the remarkably small length scale obtained from the logarithmic factor aligns precisely with our measurements and our interpretation of them as disturbances due to thermal capillary waves.

As an additional test that the breakup length is an intrinsic property of the liquid, we study actively perturbed jets. If the disturbance leading to breakup is intrinsic, the jet length should go continuously to its unperturbed length as the amplitude of the applied perturbation is decreased to zero. On the other hand, one would anticipate a nonzero minimum amplitude for inducing breakup if the noise is of external origin. We examine this by applying a fixed sinusoidal disturbance of \SI{5.4}{\kilo \hertz} (close to the Rayleigh frequency), using an immersed piezoelectric transducer. By increasing the amplitude (voltage) of the perturbation, the jet length is continuously shortened, corresponding to a smaller logarithmic pre-factor $\ln{\left(R/\delta \right)}$ in \eqref{eq:breaktime}. Figure \ref{fig:exponental_fit} (b) shows this pre-factor as a function of the transducer voltage, where $\delta$ is obtained through exponential fits as in panel (a). The jet length goes monotonically to its unperturbed length as the voltage decreases to zero, without a plateau, demonstrating that the active perturbation reaches the level of irremovable perturbations $\delta_0$ at a near-zero voltage level. This means that $\delta_{0}$ must be extremely small. In fact, our transducer changes $\delta$ by a factor $e^2 \simeq 8$ by increasing the amplitude; since the original perturbation is at the \AA-level, the perturbation imposed by the transducer is nanometric and likely still in the linear regime.

The natural breakup process of an `unperturbed' jet, therefore, remains to be understood. Even in the absence of any environmental disturbances the jet's length cannot be extended indefinitely, implying that there is always some residual disturbance $\delta_{0}$. Assuming a noisy origin of this residual disturbance, one would expect that on average the jet should break up at the fastest growing mode, i.e. the Rayleigh frequency. The natural breakup length $L$ thus corresponds to equation \eqref{eq:breaktime} calculated at the Rayleigh wavelength and with $\delta_k=\delta_0$. This yields the familiar scaling relation for the breakup length of an unperturbed jet \citep{weber1931zerfall,phinney1972stability}, which we write in terms of the diameter $D = 2 R$
\begin{equation}
\label{eq:scaling}
\frac{L}{D} = \ln{\left(\frac{D}{2\delta_0}\right)} \; \text{We}^{1/2} \, \left(1+3Z\right) \; ,
\end{equation}
where $\text{We}= \rho D V^2/\gamma \,$ is the Weber number and $Z$ is the previously defined Ohnesorge number.  Predicting the natural breakup length of a jet thus requires knowing $\delta_0$, which sets the coefficient of equation \eqref{eq:breaktime} and scaling law \eqref{eq:scaling}, $\; \ln{\left(R/\delta\right)}=\ln{\left(D/2\delta\right)} \;$.

To estimate the initial perturbation we turn to capillary wave theory (CWT). Because the spectrum of thermal capillary waves corresponds to the Fourier transform of the thermally roughened surface, we can define $\delta_0$ as the average effective amplitude of thermal capillary modes at the Rayleigh wavelength.
A liquid cylinder of area $A$ and temperature $T$ has an effective capillary wave spectrum
\begin{equation}
    \delta_{\text{eff}}(k) = R \; \sqrt{\frac{k_{B}T}{\gamma A}\frac{1}{\lvert \left(k R \right)^2 -1 \rvert }} \quad ,
    \label{eq:effective_spectrum}
\end{equation}
where $k_B$ is the Boltzmann constant. Equation (\ref{eq:effective_spectrum}) can be derived using the Equipartion Theorem and for stable modes ($\lvert kR \rvert \geq1$), $\delta_{\text{eff}}$ represents the average amplitude of sinusoidal thermal perturbations at thermal equilibrium (the 'static spectrum'; \citep{Zhang_Sprittles_Lockerby_2021}). Now the excess surface energy $\delta E$ of an infinitesimally small sinusoidal perturbation on a cylinder of area $A$ is given by $\delta E = \gamma \, A \, \left( k^2 - 1/R^2 \right)\, \delta_k^2 /2\,$, which equals $k_B T/2$ at thermal equilibrium. Unstable modes that are amplified in time through \eqref{eq:Rayleigh} strictly speaking are not in equilibrium, making \eqref{eq:effective_spectrum} an effective spectrum with an associated average amplitude $\delta_{\text{eff}}(k)$. 

The next step is to choose the appropriate value of the area $A$. The breakup process should be dominated by local perturbations, since equation \eqref{eq:effective_spectrum} requires the smallest possible (local) area $\, A_\text{min} \,$ to produce the largest possible perturbation. Capillary waves are normal modes that can only exist on the jet at a minimum length $L_0$ of the order of a wavelength. This sets $L_0=\lambda_R \simeq \sqrt{2} \pi D \,$ and $\, A_\text{min}=\pi D L_0 \simeq  \sqrt{2} \pi^2 D^2 \,$, which when introduced into \eqref{eq:effective_spectrum} together with the resonance condition $k = k_R$ yields
\begin{equation}
   \delta_0 = \sqrt{\frac{k_B \, T }{\sqrt{8}\, \pi^2 \, \gamma }} \quad .
   \label{eq:delta0}
\end{equation}
For water around room temperature ($\gamma=73 \; \text{mN}\,\text{m}^{-1}$, $\, T \sim 300 \, K \, $) $\delta_0 = 5 \cdot 10^{-11} \, \text{m} \sim 10^{-10} \, \text{m} \, $, which is consistent with our earlier observations. Note that this predicts a constant $\delta_0$ for any jet size, which means that the pre-factor $\, \ln{\left( R/\delta_0 \right)} \, $ of equation \eqref{eq:breaktime} and the scaling law \eqref{eq:scaling} depends logarithmically on the radius $R \, $. Finally, filling in equation \eqref{eq:delta0} into \eqref{eq:scaling}, gives for the jet breakup length
\begin{equation}
   \frac{L}{D}  =  \frac{1}{2} \, \ln{\left( \frac{\pi^2 \, \gamma \, D^2}{\sqrt{2} \; k_B \, T} \right)} \; \text{We}^{1/2} \, \left( 1+3 \, Z \right) \; .
   \label{eq:final_scaling}
\end{equation}
Figure \ref{fig:plot_jet_lengths} shows the predicted breakup length \eqref{eq:final_scaling} against the measured breakup length for various experimental conditions. The breakup length taken from MD simulations, at experimentally unattainable jet sizes, is also included \citep{moseler2000formation}. The fit shows excellent agreement over four orders of magnitude in measured breakup lengths (seven including MD simulations), showing that capillary jets are limited by thermal fluctuations.

Even though we propose a thermally driven model, the dependence on $T$ is more difficult to probe: for water the accessible range in absolute temperature is too small to significantly change the logarithmic factor. For `cold' liquids such as liquid nitrogen, their much lower surface tension compensates for the lower temperature in \eqref{eq:final_scaling}. Further work with exotic jets at atypical temperatures could prove invaluable, such as Helium jets \citep{Helium} or systems with ultralow interfacial tension \citep{hennequin2006drop}. Note further that the Ohnesorge number is often very small and can be neglected, so that the Rayleigh wavelength can be approximated as $\lambda_R = \pi D \, \sqrt{2 \left( 1+3 \, Z \right)} \simeq \sqrt{2}\pi D \,$.

In conclusion, we challenge the long-standing belief that external noise or nozzle imperfections are the primary causes of laminar jet breakup. While strong external disturbances can affect the jet length, our findings show that under typical experimental conditions external influences become negligible. In the laminar regime, jet breakup lengths show no significant variation across diverse liquids and nozzle geometries, suggesting a unified breakup mechanism independent of nozzle quality or flow profile. In fact, intrinsic thermal capillary waves set an unavoidable background of effective disturbances at unstable Rayleigh-Plateau frequencies, and our results suggest that other modes e.g. induced by nozzle imperfections grow more slowly than the Rayleigh frequency. Extensive testing and analysis demonstrate that these small thermal fluctuations are the fundamental cause of jet breakup. Our thermally induced breakup model accurately predicts jet breakup lengths over more than seven orders of magnitude, from nanojets to macroscopic jets, without adjustable parameters. These findings have important implications for various applications involving liquid jets, from inkjet printing to drug delivery. By understanding that thermal capillary waves are the fundamental source of jet instability, we can better predict and control jet behavior across a wide range of conditions.

\section{acknowledgments}
This research has been funded by the Dutch Research Council NWO, IPP grant “Innovative Nanotech Sprays”, ENPPS.IPP.019.001, and supported by Fundação para a Ciência e Tecnologia (FCT). 

%

\end{document}